\documentclass[a4paper,11pt]{article}

\usepackage{amsmath}
\usepackage{amssymb}
\usepackage{color}
\usepackage[dvips]{graphicx}
\usepackage{cite}

\makeatletter
\@addtoreset{equation}{section}
\renewcommand{\theequation}{\thesection.\@arabic\c@equation}
\makeatother

\def \be {\begin{equation}}
\def \ee {\end{equation}}
\def \ba {\begin{array}}
\def \ea {\end{array}}
\def \bea{\begin{eqnarray}}
\def \eea{\end{eqnarray}}

\def \a {\alpha}
\def \b {\beta}
\def \g {\gamma}
\def \G {\Gamma}
\def \d {\delta}
\def \D {\Delta}
\def \e {\epsilon}

\def \m {\mu}
\def \n {\nu}

\def \l {\lambda}
\def \L {\Lambda}
\def \s {\sigma}
\def \S {\Sigma}
\def \r {\rho}
\def \o {\omega}
\def \O {\Omega}

\def \t {\tau}

\def \p {\partial}
\def \f {\frac}
\def \na {\nabla}

\def \nn {\nonumber}
\def \ma {\mathcal}

\def \lt {\left}
\def \rt {\right}

\def \ra {\rightarrow}
\def \sr {\sqrt}
\def \td {\tilde}
\def \hs {\hspace}
\def \pp {\propto}
\def \inf {\infty}

\title{\textbf{Notes on Self-dual Warped dS$_3$ Spacetime}}
\author{
Bin Chen$^{1,2}$\footnote{bchen01@pku.edu.cn}
\hs{3ex}
Jia-ju Zhang$^1$\footnote{jjzhang@pku.edu.cn}
}
\date{}

\begin{document}

\maketitle

\begin{center}
\vspace{5mm}
{\it
$^{1}$Department of Physics, and State Key Laboratory of Nuclear Physics and Technology,\\
Peking University, Beijing 100871, P.R. China\\
\vspace{2mm}
$^{2}$Center for High Energy Physics, Peking University, Beijing 100871, P.R. China\\
}
\vspace{10mm}
\end{center}

\begin{abstract}

In this paper, we investigate the warped dS/CFT correspondence  of the self-dual warped dS$_3$ spacetime, which is a solution of three-dimensional topologically massive gravity (TMG) with a positive cosmological constant. We discuss its thermodynamics at two cosmological horizons. We propose that the self-dual warped dS$_3$  spacetime is holographically dual to a chiral conformal field theory.   We analytically calculate the quasinormal modes of scalar, vector, tensor, and spinor perturbations under the self-dual warped dS$_3$ spacetime.  It turns out that the frequencies of the quasinormal modes could be identified with the poles in the thermal boundary-boundary correlators.

\end{abstract}

\baselineskip 18pt

\thispagestyle{empty}

\newpage

\section{Introduction}

 Three dimensional topologically massive gravity(TMG)\cite{TMG} provides an interesting modification of Einstein's general relativity. Besides the Einstein-Hilbert term and  the cosmological constant $\L$, the action of TMG theory includes also a gravitational Chern-Simons term
\bea \label{z5}
&&I_{TMG}=I_{EH}+I_{CS}  \nn\\
&&\phantom{I_{TMG}}
=\f{1}{16\pi G}\int d^3x \sr{-g} \lt[ R-2\L  +\f{1}{2\m}\e^{\l\m\n}\G^\r_{\l\s}
                             \lt( \p_\m\G^\s_{\r\n} +\f{2}{3}\G^\s_{\m\t}\G^\t_{\n\r} \rt) \rt].
\eea
Unlike the usual three-dimensional pure gravity which has no local degree of freedom, there is generically a massive propagating degree of freedom in TMG. The equations of motion in TMG is a third order differential equation, involving the Cotton tensor. As a result, the usual solutions of the  Einstein gravity with/without a cosmological constant, which must be of a Einstein metric, are also the solutions of the TMG theory. Moreover there are new solutions which has nontrivial Cotton tensor. For the TMG with a cosmological constant, these solutions include the spacelike, timelike, null warped spacetimes, and also regular black hole solutions.

In the past few years,  3D TMG with a negative cosmological constant has brought us many surprises, especially in the context of AdS/CFT correspondence.
Inspired by the AdS$_3$/CFT$_2$ correspondence\cite{BrownHenneaux,Strominger:1997eq}, it was proposed in \cite{Anninos:2008fx} that the quantum gravity in the warped AdS$_3$ spacetime is dual to the CFT with the central charges
\be\label{central}
c_L=\f{4\n\ell}{G(\n^2+3)},\hs{3ex} c_R=\f{(5\n^2+3)\ell}{G\n(\n^2+3)},
 \ee
 taking the cosmological constant $\L=-1/\ell^2$ and the parameters $\m=3\n/\ell$. This warped AdS/CFT correspondence was set up by
 analyzing the thermodynamic properties of the spacelike stretched and null warped AdS$_3$ black holes. Later on, the central charges were confirmed by studying the asymptotic symmetries of the spacelike stretched warped AdS$_3$ spacetime \cite{Compere:2008cv,Compere:2009zj,Blagojevic:2009ek,Henneaux2011}. The warped AdS/CFT correspondence was further confirmed by the computation of the quasinormal modes and real-time correlator in the spacelike stretched and null warped AdS$_3$ black holes \cite{Chen:2009rf,Chen:2009hg,Chen:2009cg}.

 Moreover, the warped AdS/CFT correspondence was proposed for the self-dual warped AdS$_3$ spacetime\cite{Chen:2010qm}. This spacetime takes a similar form to the near-NHEK geometry
\cite{Bredberg:2009pv,Chen:2010ni} and is asymptotic to the spacelike warped AdS$_3$ without any coordinate transformation. It was shown that for appropriate boundary conditions, the asymptotic symmetry generators form one copy of the Virasoro algebra with central charge $c_L$ in (\ref{central}), with which the
application of the Cardy formula  precisely reproduces the Bekenstein-Hawking entropy. The warped AdS/CFT correspondence in this casee is a chiral one, very different from the one suggested in \cite{Anninos:2008fx}. The boundary conditions proposed are different from those suggested in \cite{Compere:2008cv,Compere:2009zj}. It is remarkable that  the left central charge arises
naturally through a centrally extended Virasoro algebra which is an enhancement of the $U(1)_L$ isometry, instead of a Sugawara-type
procedure from a current algebra as in \cite{Blagojevic:2009ek}.

% There is also the self-dual warped AdS$_3$ black hole as the solution of the TMG theory, which provides another confirmation of the warped AdS/CFT correspondence \cite{Chen:2010qm,Li:2010sv}.\footnote{}

Similar to AdS/CFT correspondence, there are also the proposed dS/CFT correspondences \cite{Park,Strominger:2001pn}. It was conjectured that there is a dual Euclidean CFT living at the future timelike infinity $\ma I^+$ of a de-Sitter spacetime. Moreover, the warped AdS/CFT correspondence has also been generalized to the warped de-Sitter case\cite{Anninos:2009jt}. The holography for the warped dS$_3$ spacetimes were investigated in \cite{Anninos:2011vd}. In the paper,  the thermodynamics and the asymptotic structure of a family of warped dS$_3$ geometries were studied, and it was conjectured that quantum gravity in asymptotically warped dS$_3$ is holographically dual to a two-dimensional conformal field theory living at $\ma I^+$. The investigation of the quasinormal modes and hidden conformal symmetry in \cite{Chen:2011dc} supports the warped dS/CFT correspondence.

In this paper we will provide another warped dS/CFT correspondence for the self-dual warped dS$_3$ spacetime. The correspondence has firstly been proposed in \cite{Anninos:2009jt} for the warped de-Sitter spacetime. It was conjectured that the self-dual warped dS$_3$ spacetime is dual to a chiral CFT with nonvanishing left central charge $c_L=\frac{4\nu \ell}{G(3-\nu^2)}$. The picture is quite similar to the one for the self-dual warped AdS$_3$ spacetime. In the next section, we consider a more general self-dual warped dS$_3$ spacetime, which could be  related to the self-dual warped AdS$_3$ spacetime in \cite{Chen:2010qm} by analytic continuation. We discuss its geometric and thermodynamical properties carefully and clarify its relation with the warped dS$_3$ spacetime. Then we propose the CFT dual for this class of spacetime.

  In Sec. 3 we solve the scalar equation in the warped dS$_3$ black hole background. We discuss the solutions in the physical region between two cosmological horizons and the region outside the horizon. From the late-time behavior, we read the thermal boundary-boundary correlators. We also find the hidden conformal symmetry in the scalar equation, which leads to the same dual temperatures as the ones got from Frolov-Thorne vacuum. In Sect. 4 we compute the quasinormal modes of various perturbations by imposing proper boundary conditions at the cosmological horizons. We end with some discussions in Sec. 5.

\section{Self-dual warped dS$_3$ spacetime}

The metric of the warped de Sitter space can be written in the static patch as
\be \label{z6}
ds^2=\f{\ell^2}{3-\n^2} \lt( -(1-r^2)d\t^2+\f{dr^2}{1-r^2}+\f{4\n^2}{3-\n^2} (du+rd\t)^2 \rt),
\ee
Here $(\t,u)\in  R^2$, $r^2<1$ for the patches within the cosmological horizons and $r^2>1$ for the patches between the horizons and the future and past timelike infinity $\ma I^\pm$. It is the vacuum solution of TMG action (\ref{z5}) with the positive cosmological constant $\L=1/\ell^2$, and satisfies the equation of motion
\be
G_{\m\n}+\f{1}{\ell^2}g_{\m\n}+\f{1}{\m}C_{\m\n}=0,
\ee
where $G_{\m\n}$ is the Einstein tensor and $C_{\m\n}$ is the Cotton tensor
\be
C_{\m\n}=\e_{\m}^{\phantom{\m}\a\b}\na_\a \lt( R_{\b\n}-\f{1}{4}g_{\b\n}R \rt).
\ee
We have taken the convention that $\e^{\t ru}=\f{-1}{\sr{-g}}$, $\n>0$, and thus $\m=-\f{3\n}{\ell}$. The entropy of this spacetime is proportional to the integral
over $u$, which is divergent. One may requires $u$ to be of period $2\pi$ such that the spacetime becomes self-dual with a finite entropy. It was proposed in \cite{Anninos:2009jt} that the spacetime is holographically dual to a CFT with a nonvanishing $c_L=\frac{4\nu \ell}{G(3-\nu^2)}$ and a temperature $T_L=1/2\pi \ell$. The match of the microscopic CFT entropy with the macroscopic entropy  supports this conjecture.

There is a more general class of self-dual warped dS$_3$ spacetime in TMG theory, whose metric is of the form
\be \label{e1}
ds^2=\f{\ell^2}{3-\n^2} \lt[ -(r-r_b)(r_c-r)dt^2+\f{dr^2}{(r-r_b)(r_c-r)}
                             +\f{4\n^2}{3-\n^2} \lt( \a d\phi+ (r-\f{r_b+r_c}{2})dt \rt)^2 \rt],
\ee
with $r_b < r_c$ being two cosmological horizons. It is free of curvature singularity and regular everywhere, even though there exist two Killing horizons corresponding to the Killing vector of time translations. As there is a freedom in translating $r$, only the
difference of $r_c-r_b$ is meaningful.

Actually, the spacetime (\ref{e1})  could be obtained  from its AdS$_3$ cousin discussed in \cite{Chen:2010qm} by analytic continuation
\be
\ell \to i \ell, \hs{3ex} \n \to -i \n.\label{ana}
\ee
It was shown that the self-dual warped AdS$_3$ spacetime is dual to a chiral CFT with
\be
c_L = \frac{4\nu \ell}{G(\nu^2+3)}, \hs{3ex}T_L={\frac{\alpha}{2\pi \ell}}.\label{centralA}
\ee
In this case, there is actually a nonvanishing right-moving temperature $T_R = \frac{r_c - r_b}{4\pi \ell}$. The situation is reminiscent of AdS$_2$ black hole. As discussed in detail in \cite{Maldacena:1998uz,Spradlin:1999bn}, the observers moving along the worldlines of fixed $r$ will detect a thermal bath of the particles at temperature $T_R$
in the global Hartle-Hawking vacuum, essentially similar to the case that Rindler observers detect the radiations in the Minkowski vacuum.

It is also remarkable that the spacetime (\ref{e1}) could be obtained as the Nariai limit of the warped dS$_3$ black hole\cite{Anninos:2011vd}.

On the other hand,  the spacetime (\ref{e1}) has regular thermodynamic behavior, satisfying the first law of thermodynamics.
The Hawking temperatures and angular velocities of the two horizons are
\bea
&& T_b=T_c=\f{r_c-r_b}{4\pi\ell},  \nn\\
&& \O_b=-\O_c=\f{r_c-r_b}{2\a\ell}.
\eea
Using the formalism developed in \cite{Solodukhin:2005ah,Tachikawa:2006sz}, the entropy  at the horizon $r=r_h$ has two contribution, one  from the Einstein-Hilbert term and the other from the Chern-Simons term \cite{Chen:2011dc},
\bea
&&S_{EH}=\lt.\f{\pi \sr{g_{\phi\phi}}}{2G}\rt|_{r=r_h}, \nn\\
&&S_{CS}=\lt.\f{\pi g_{\phi\phi}\p_r N^\phi}{4\m G \sr{N^2 g_{rr}}}\rt|_{r=r_h}.
\eea
It turns out that the entropies of the two horizons are the same,
\be \label{z7}
S_b=S_c=\f{2\pi\a\n\ell}{3G(3-\n^2)}.
\ee
The Abbott-Deser-Tekin (ADT) mass and angular momentum are defined as surface integrals corresponding to the conserved charges associated to the asymptotic  Killing vectors $\p_t$ and $\p_\phi$ \cite{Abbott:1982jh,Abbott:1982ff,Deser:2002rt,Deser:2002jk,Deser:2003vh}, and were generalized to  arbitrary backgrounds in the TMG theory in \cite{Bouchareb:2007yx}. To the self-dual warped dS$_3$ spacetime, they are
\be
\ma M_b=\ma M_c=0, ~~~ \ma J_b=-\ma J_c=\f{(\a^2-1)\n\ell}{6G(3-\n^2)},
\ee
where the signs of angular momenta are chosen for the first law of thermodynamics to be satisfied
\bea
&& d\ma M_b=T_b d S_b +\O_b d\ma J_b,  \nn\\
&& d\ma M_c=T_c d S_c +\O_c d\ma J_c.
\eea
We have chosen the vacuum to be $\a=1$, because then the spacetime metric (\ref{e1}) and the metric of warped dS$_3$ spacetime (\ref{z6}) have the same asymptotic behavior. Since now the energy vanishes identically, there is no problem about the positivity of the energy.

The left and right moving temperatures  of the dual CFT can be got from the Frolov-Thorne vacuum\cite{Frolov:1989jh}
\be \label{e7}
T_L=\f{\a}{2\pi \ell}, ~~~ T_R=\f{r_c -r_b}{4\pi \ell}.
\ee
It is clear that the conserved charges and the entropy of the spacetime is independent of $r_c-r_b$ and therefore the right-moving temperature, suggesting that the dual CFT is chiral, even though in the dual CFT both left-moving and right-moving temperatures are nonvanishing. The Cardy formula
\be
S_{CFT}=\f{\pi^2 \ell}{3}(c_L T_L + c_R T_R)
\ee
 reproduces the entropy of the spacetime provided
\be
c_L=\f{4\n\ell}{G(3-\n^2)}, ~~~ c_R=0.\label{central2}
\ee
Note that the above central charges  are related to the ones (\ref{centralA}) for
self-dual warped AdS$_3$ by analytic continuation (\ref{ana}).

In short, we think that the spacetime (\ref{e1}) is holographically dual to a chiral CFT
with the central charges (\ref{central2}) and the temperatures (\ref{e7}). In the following, we will discuss the two-point functions and hidden conformal symmetry to support this conjecture.

\section{Scalar wave equation}

The calculation of the scalar equation is parallel to that in \cite{Chen:2011dc}. We first solve the scalar equation scattering off the self-dual warped dS$_3$ spacetime in the physical regions $r_b<r<r_c$ between two cosmological horizons. Then we discuss the solution outside the horizon with $r>r_c$, and moreover obtain the boundary-boundary thermal correlator from the late-time behavior of the scalar field in the region $r>r_c$.

\subsection{In the region $r_b<r<r_c$}

The equation of the massive scalar $\Phi=e^{-i\o t+ik\phi}R(r)$ in the background (\ref{e1}) is
\be \label{z4}
\na_\m \na^\m \Phi=\td \D \Phi=m^2 \phi,
\ee
where we have defined the operator
\be
\td \D=\f{1}{\sr{-g}}\p_\m \sr{-g} g^{\m\n} \p_\n.
\ee
The scalar equation can be calculated explicitly as
\bea \label{e2}
&&\p_r(r-r_b)(r_c-r)\p_r R(r)
+\f{[2\a\o-(r_c-r_b)k]^2}{4\a^2(r_c-r_b)(r-r_b)} R(r)
+\f{[2\a\o+(r_c-r_b)k]^2}{4\a^2(r_c-r_b)(r_c-r)} R(r) \nn\\
&&=\lt( \f{m^2\ell^2}{3-\n^2}+\f{3(1+\n^2)k^2}{4\a^2\n^2} \rt) R(r).
\eea
We solve the scalar equation in the region $r_b<r<r_c$. With the new coordinate
\be
z=-\f{r-r_b}{r_c-r},
\ee
we have two independent solutions
\bea
&&R_1=(-z)^{\a_s}(1-z)^{\b_s} F(a,b,c;z),  \nn\\
&&R_2=(-z)^{-\a_s}(1-z)^{\b_s} F(a-c+1,b-c+1,2-c;z),
\eea
where $F(a,b,c;z)$ is the hypergeometric function and
\bea \label{e3}
\a_s&=&-i\f{2\a\o-(r_c-r_b)k}{2\a(r_c-r_b)},    \nn\\
\b_s&=&\f{1}{2}+\sr{\f{1}{4}-\f{m^2\ell^2}{3-\n^2}-\f{3(1+\n^2)k^2}{4\a^2\n^2}},  \nn\\
\g_s&=&-i\f{2\a\o+(r_c-r_b)k}{2\a(r_c-r_b)},  \nn\\
a&=&\a_s+\b_s-\g_s=\b_s+i\f{k}{\a},  \nn\\
b&=&\a_s+\b_s+\g_s=\b_s-i\f{2\o}{r_c-r_b},  \nn\\
c&=&1+2\a_s.
\eea

For physical solution we impose the outgoing boundary condition at the cosmological horizons. The boundary condition at the horizon $r_b$ requires us to choose the first solution $R_1$ and discard the second one $R_2$. When $r$ goes from $r_b$ to $r_c$, $z$ goes from 0 to $-\inf$. When $|z| \ra \inf$,
\be
F(a,b,c;z) \sim \f{\G(c)\G(b-a)}{\G(c-a)\G(b)}(-z)^{-a}+\f{\G(c)\G(a-b)}{\G(c-b)\G(a)}(-z)^{-b},
~~  |\arg(-z)|<\pi,
\ee
and then when $r \ra r_c $ we have
\be
R_1 \sim A_{in} (-z)^{\g_s}+  A_{out} (-z)^{-\g_s},
\ee
with
\be
A_{in}=\f{\G(c)\G(b-a)}{\G(c-a)\G(b)}, ~~~ A_{out}= \f{\G(c)\G(a-b)}{\G(c-b)\G(a)}.
\ee
The outgoing boundary condition at the cosmological horizon $r_c$ then requires that  the first term of the above equation must vanish
\be \label{e10}
A_{in}=0.
\ee

\subsection{In the region $r>r_c$}

In the region $r>r_c$, the scalar equation (\ref{e2}) can be solved in terms of the new variable
\be
x=\f{r-r_c}{r-r_b},
\ee
and there are two independent solutions
\bea
&&\td R_1=x^{-\g_s}(1-x)^{\b_s}F(\td a,\td b,\td c;x),\nn\\
&&\td R_2=x^{\g_s}(1-x)^{\b_s}F(\td a-\td c+1,\td b-\td c+1,2-\td c;x),
\eea
with $\a_s, ~ \b_s, ~ \g_s$ being the same as (\ref{e3}) and
\bea
&&\td a=a=\b_s+\a_s-\g_s=\b_s+i\f{k}{\a},  \nn\\
&&\td b=\b_s-\a_s-\g_s=\b_s+i\f{2\o}{r_c-r_b},  \nn\\
&&\td c=1-2\g_s.
\eea

At the cosmological horizon $r\ra r_c$, we choose the boundary condition that the ingoing mode $\td R_2$ vanishes and the outgoing mode $\td R_1$ coincides with the outgoing mode $R_1$ in the region $r<r_c$ calculated in the last subsection with a proper overall coefficient\cite{Anninos:2010gh}. Outside  the cosmological horizon $r>r_c$, $r$ behaves as the timelike coordinate. As $r\ra\infty$,  $x\ra1,~1-x\ra r^{-1}$, we have the late-time behavior
\be \label{e4}
\td R_1 \sim C_1 r^{\b_s-1}+ C_2 r^{-\b_s},
\ee
with
\be
C_1=\f{\G(2\b_s-1)\G(\td c)}{\G(\td a)\G(\td b)},
~~~ C_2=\f{\G(1-2\b_s)\G(\td c)}{\G(\td c-\td a)\G(\td c-\td b)}.
\ee
We suppose that the scalar solution has the asymptotic behavior $R \sim r^{-h}$ so that we may identify the conformal weights
\bea
&&h_L=h_R=\b_s=\f{1}{2}+\sr{\f{1}{4} -\f{m^2\ell^2}{3-\n^2} -\f{3(1+\n^2)k^2}{4\a^2\n^2} }, ~~~ \textrm{or} \nn\\
&&h_L=h_R=1-\b_s=\f{1}{2}-\sr{\f{1}{4} -\f{m^2\ell^2}{3-\n^2} -\f{3(1+\n^2)k^2}{4\a^2\n^2} }.
\eea
The study from the hidden conformal symmetry will confirm this identification.

\subsection{Two-point functions}

When $r \ra \inf$ the above warped dS$_3$ spacetime metric (\ref{z6}) becomes
\be
ds^2=\f{\ell^2}{3-\n^2} \lt( r^2 d\t^2 -\f{dr^2}{r^2}+\f{4\n^2}{3-\n^2} (du+rd\t)^2 \rt).
\ee
And in the region $r\ra\inf$, the background  (\ref{e1}) behaves as
\be
ds^2=\f{\ell^2}{3-\n^2} \lt( r^2 dt^2 -\f{dr^2}{r^2}+\f{4\n^2}{3-\n^2} (\a d\phi+rdt)^2 \rt).
\ee
With $u=\a\phi$, the two asymptotic geometries are the same. Let $\tilde k$ be the quantum number along $u$, then we have  the quantum number identifications
\be \label{z3}
\td k=\f{k}{\a}.
\ee
%As what was done in \cite{Chen:2009hg,Chen:2011dc}, we have to use the new quantum number $\td k$, in the setup of the warped dS/CFT correspondence.

The coefficients $\td a, ~ \td b$ in the last subsection can be expressed in terms of the CFT parameters: the conformal weights $(h_L,h_R)$, and the frequencies $(\o_L,\o_R)$
\be
\td a=h_L + i\f{\o_L}{2\pi T_L},~~~
\td b=h_R + i\f{\o_R}{2\pi T_R},
\ee
and so
\be
i\g=\f{\o_L}{4\pi T_L}+\f{\o_R}{4\pi T_R},
\ee
with
\bea
&& h_L=h_R=\b_s \nn\\
&& \o_L=k, ~~~ \o_R=\o.
\eea

Formally, we may define the boundary-boundary thermal correlator  from $C_1, ~ C_2$ in (\ref{e4}) as what was done in \cite{Anninos:2010gh}
\bea
&&G_R \sim \f{C_2}{C_1} \pp \f{\G(\td a)\G(\td b)}{\G(\td c-\td a)\G(\td c-\td b)} \nn\\
&&  \phantom{G_R}  \pp \sin \lt(\pi h_L - i \f{\o_L}{2T_L} \rt)
                          \sin \lt(\pi h_R - i \f{\o_R }{2T_R} \rt)  \nn\\
&&\phantom{G_R\pp} \times \G \lt( h_L-i \f{\o_L }{2\pi T_L} \rt)
                            \G \lt( h_L+i \f{\o_L }{2\pi T_L} \rt)  \nn\\
&&\phantom{G_R\pp} \times\G \lt( h_R-i \f{\o_R}{2\pi T_R} \rt)
                            \G \lt( h_R+i \f{\o_R}{2\pi T_R} \rt).
\eea
As argued in \cite{Anninos:2011vd}, the possible way to understand it is to do double Wick rotation such that the radial direction becomes really spacelike and the translation along $t$ becomes timelike. Then the above correlator could be considered as a retarded Green's function. The above correlators is reminiscent of the Euclidean CFT correlators. Recall that in a 2D CFT, by conformal symmetry the Euclidean correlator takes the form\cite{Cardy:1984bb}
\bea \label{e5}
&&G_E \sim T_L^{2h_L-1}T_R^{2h_R-1}
         e^{ i \o_{L,E}/2T_L }e^{ i \o_{R,E}/2T_R }\nn\\
&&\phantom{G_E \sim}        \times \G \lt( h_L- \f{\o_{L,E}}{2\pi T_L} \rt)
                                   \G \lt( h_L+ \f{\o_{L,E}}{2\pi T_L} \rt) \nn\\
&&\phantom{G_E \sim}        \times \G \lt( h_R- \f{\o_{R,E}}{2\pi T_R} \rt)
                                   \G \lt( h_R+ \f{\o_{R,E}}{2\pi T_R} \rt),
\eea
with the Euclidean frequencies
\be
\o_{L,E}=i\o_L, ~~~
\o_{R,E}=i\o_R.
\ee

As the conformal weight could be smaller than 1, we may alternate the roles of the source and the response and define the boundary-boundary thermal correlator as
\be
G_R \sim \f{C_1}{C_2},
\ee
and then we can get the same results as before with different conformal weights
\be
h_L=h_R=1-\b_s.
\ee
In both cases, we can read the poles in the correlator (\ref{e5}) as
\be \label{z2}
\o_R=-i2\pi T_R (n+h_R).
\ee

From the first law of black hole thermodynamics and its CFT counterpart
\be
\d S_{CFT}=\f{\d E_L}{T_L}+\f{\d E_R}{T_R},
\ee
we get the conjugate charges $(\d E_L,\d E_R)$ as
\bea
&&\d E_L=\d J_c, \nn\\
&&\d E_R=\d M_c.
\eea
The identifications of parameters are $\d M_c=\o,~\d J_c=k$, and then we have
\bea
&&\o_L=\d E_L \lt( \d M_c=\o;\d J_c=k\rt), \nn\\
&&\o_R=\d E_R \lt( \d M_c=\o;\d J_c=k \rt).
\eea

\subsection{Hidden conformal symmetry}

The hidden conformal symmetry  is an useful concept in finding the holographic dual of the black hole.
The hidden conformal symmetry of black holes was first proposed in the low frequency scattering off generic nonextreme Kerr black hole\cite{Castro:2010fd}. It was found that the scalar equation of motion in the near region could be written in terms of $SL(2,R)$ Casimir, with which the temperatures of the dual CFT can be read off directly.   For three-dimensional black holes, the hidden conformal symmetry could often be defined in the whole spacetime rather than just the near region. This makes the study of the hidden conformal symmetry in three-dimensional black holes more interesting and tractable. In \cite{Chen:2010ik}, it was shown that for general tensor fields, their equations of motion in three-dimensional BTZ and warped black holes could be written in terms of Lie-induced SL(2,R) Casimir as well. This fact indicates that the hidden conformal symmetry is an intrinsic property of the black hole.

To the self-dual warped dS$_3$ spacetime (\ref{e4}) we analyze the hidden conformal symmetry in the physical region between two cosmological horizons $r_b<r<r_c$. In this region,
 we may define the conformal coordinates
\bea
&&\o^+=\sr{\f{r_c-r}{r-r_b}}e^{2\pi T_R t+2n_R \phi},  \nn\\
&&\o^-=\sr{\f{r_c-r}{r-r_b}}e^{2n_L t + 2\pi T_L\phi},  \nn\\
&&y=\sr{\f{r_c-r_b}{r-r_b}}e^{(\pi T_R+n_L)t+(\pi T_L+n_R)\phi},
\eea
Note that they are different from those in \cite{Chen:2011dc}, but similar to the ones for  the self-dual AdS$_3$ black holes\cite{Li:2010zr}.  The vector fields $(V_0,V_\pm)$ and $(\td V_0, \td V_\pm)$ could be locally defined as
\bea
&&V_1=\p_+, \nn\\
&&V_0=\o^+\p_++\frac{1}{2}y\p_y, \nn\\
&&V_{-1}=\o^{+2}\p_++\o^+y\p_y+y^2\p_-,
\eea
\bea
&&\tilde V_1=\p_- \nn\\
&&\tilde V_0=\o^-\p_-+\frac{1}{2}y\p_y \nn\\
&&\tilde V_{-1}=\o^{-2}\p_-+\o^-y\p_y+y^2\p_+.
\eea
These vector fields obey the $SL(2,R)$ Lie algebra
\be
[V_0, V_{\pm 1}]=\mp V_{\pm 1},\hs{5ex} [V_{-1},V_1]=-2 V_0,
\ee
and similarly for $(\tilde V_0, \tilde V_{\pm 1})$.
Written in the coordinates $(t,r,\phi)$, these vector fields are
\bea
&&V_1=e^{-2\pi T_R t-2 n_R \phi} \lt( -\sr{\D}\p_r   -\f{\pi T_L \D'-n_R(r_c-r_b)}{4A\sr{\D}}\p_t
                                     +\f{n_L \D'-\pi T_R(r_c-r_b)}{4 A\sr{\D}}\p_\phi \rt),  \nn\\
&&V_0=\f{\pi T_L \p_t-n_L\p_\phi}{2A}, \nn\\
&&V_{-1}=e^{2\pi T_R t+2 n_R \phi} \lt( -\sr{\D}\p_r   +\f{\pi T_L \D'-n_R(r_c-r_b)}{4A\sr{\D}}\p_t
                                     -\f{n_L \D'-\pi T_R(r_c-r_b)}{4 A\sr{\D}}\p_\phi \rt),  \nn\\
\eea
\bea
&&\td V_1=e^{-2 n_L t-2\pi T_L \phi} \lt( -\sr{\D}\p_r   +\f{n_R \D'-\pi T_L(r_c-r_b)}{4A\sr{\D}}\p_t
                                     -\f{\pi T_R \D'-n_L(r_c-r_b)}{4 A\sr{\D}}\p_\phi \rt),  \nn\\
&&\td V_0=\f{-n_R \p_t+\pi T_R\p_\phi}{2A}, \nn\\
&&\td V_{-1}=e^{2 n_L t+2\pi T_L \phi} \lt( -\sr{\D}\p_r   -\f{n_R \D'-\pi T_L(r_c-r_b)}{4A\sr{\D}}\p_t
                                     +\f{\pi T_R \D'-n_L(r_c-r_b)}{4 A\sr{\D}}\p_\phi \rt),  \nn\\
\eea
where we have defined $\D=(r-r_b)(r_c-r), ~ A=\pi^2 T_L T_R-n_L n_R.$

The quadratic Casimir is defined as
\bea
&&\ma H^2=\tilde{\ma H}^2=-V_0^2+\frac{1}{2}(V_1 V_{-1}+V_{-1}V_1) \nn\\
&&\phantom{\ma H^2}=\frac{1}{4}( y\p_y - y^2\p^2_y) +y^2 \p_+\p_-.
\eea
and in terms of $(t,r,\phi)$ coordinates it becomes
\bea
&&\ma H^2=\p_r \D \p_r
-\f{(r_c-r_b)[(\pi T_L - n_R)\p_t+(\pi T_R-n_L)\p_\phi]^2}{16 A^2 (r-r_b)} \nn\\
&&\phantom{\ma H^2=} -\f{(r_c-r_b)[(\pi T_L +n_R)\p_t-(\pi T_R+n_L)\p_\phi]^2}{16^2 A^2 (r_c-r)}.
\eea
With the scalar field being expanded as $\Phi=e^{-i\o t+ik\phi}R(r)$, the equation $\ma H^2\Phi=-K\Phi$  gives the radial equation of motion
\bea \label{e6}
&&\p_r \D \p_r R(r)
+\f{(r_c-r_b)[(\pi T_L-n_R)\o-(\pi T_R-n_L)k]^2}{16 A^2 (r-r_b)} R(r) \nn\\
&&+\f{(r_c-r_b)[(\pi T_L+n_R)\o+(\pi T_R+n_L)k]^2}{16 A^2 (r_c-r)} R(r)=-K R(r),
\eea
where $K$ is a constant.

Eq. (\ref{e6}) reproduces the scalar equation (\ref{e2}) under the identifications
\bea
&&K=-\f{m^2\ell^2}{3-\n^2}-\f{3(1+\n^2)k^2}{4\a^2\n^2},  \nn\\
&&T_L=\f{\a}{2\pi}, ~~~ T_R=\f{r_c-r_b}{4\pi}, \nn\\
&&n_L=n_R=0.
\eea
This suggests that the warped dS$_3$ spacetime does have the hidden conformal symmetry in the whole physical region. Another remarkable point is that in the above identifications we find the same dual temperatures $T_L$ and $T_R$ as the ones (\ref{e7}). This provides another support that the self-dual warped dS$_3$ spacetime could be described by a finite temperature CFT.

It is straightforward to generalize the above discussions on hidden conformal symmetry to the vector and tensor fluctuations along the line suggested in \cite{Chen:2010ik}.
The algebraic construction of the quasinormal modes based on hidden conformal symmetry in the next section indicates that the equations of motion of these fluctuations could be
written in terms of Lie-induced SL(2,R) Casimir.

\section{Quasinormal modes}

The quasinormal modes of the black holes are defined as the perturbations subject to proper boundary conditions at the black hole horizon and the infinity (the cosmological horizon for de Sitter black holes). The frequencies of the quasinormal modes have negative imaginary parts, indicating  the perturbations undergo damped oscillations. In AdS/CFT correspondence, the quasinormal modes correspond to perturbations in finite temperature CFT, and the decay of the modes corresponds to the relaxation of the fluctations to thermal equilibrium. The frequencies of the quasinormal modes agree with the poles of retarded Green's function of the dual CFT \cite{Horowitz:1999jd}. The quasinormal modes of the self-dual warped AdS$_3$ were discussed in \cite{Li:2010sv} and the ones of the warped dS$_3$ black holes were studied in \cite{Chen:2011dc}.

In this section we calculate the quasinormal modes of scalar, vector, tensor, and spinor perturbations under the self-dual warped dS$_3$ background.

\subsection{Scalar perturbation}

The constraint (\ref{e10}) requires
\be
b=-n, ~~ \textrm{or} ~~ c-a=-n,
\ee
with $n$ being a nonnegative integer. When $b=-n$, we have
\be
\o=-i2\pi T_R(n+\b),
\ee
and when $c-a=-n$, we have
\be
\o=-i2\pi T_R(n+1-\b).
\ee
Thus, we always have the frequencies of the quasinormal modes
\be \label{z1}
\o_R=-i2\pi T_R(n+h_R^\pm),
\ee
with
\bea
&& h_R^+=\b, ~~ \textrm{or} ~~ h_R^-=1-\b, ~~ \rm{i.e.} \nn\\
&& h_R^\pm=\f{1}{2} \pm \sr{\f{1}{4} -\f{m^2\ell^2}{3-\n^2} -\f{3(1+\n^2)\td k^2}{4\n^2} }.
\eea
The conformal weights are the same as those of the warped dS$_3$ black hole case \cite{Chen:2011dc}, and it is because the two geometries have the same asymptotic behaviors after the redefinition of the coordinates. So without surprise, the conformal weights of the self-dual warped dS$_3$ black hole can be transformed to those of the spacelike stretched warped black hole by analytical continuation. The same situations apply to the vector, tensor and spinor perturbations. The frequency always has a negative imaginary part and so has the right behavior of the quasinormal mode. Also it is easy to see that the frequencies  of the quasinormal modes (\ref{z1}) are the poles (\ref{z2}) of the boundary-boundary two-point function.

\subsection{Vector perturbation}

In three dimensions the vector equation can be written as the two first order equations
\be
\e_\l^{\phantom{\l}\m\n}\p_\m A_\n \pm m A_\l=0,
\ee
which leads to
\be
\td \D A_\phi= \lt( m^2 \pm \f{2\n m}{\ell} \rt) A_\phi.
\ee
Comparing the vector equation with the scalar equation (\ref{z4}), we can see that in order to get the quasinormal modes of the vector perturbation we just need to make the change $m^2 \ra m^2 \pm \f{2\n m}{\ell}$ in the scalar results. Then we get the quasinormal modes and conformal weights of the vector perturbation
\bea
&&\o_R=-i 2\pi T_R (n+h_R^\pm),  \nn\\
&&h_R^\pm=\f{1}{2} \pm \sr{\f{1}{4} - \f{m^2\ell^2 + 2\n m\ell}{3-\n^2} -\f{3(1+\n^2)\td k^2}{4\n^2} }, ~~ \textrm{or}  \nn\\
&&h_R^\pm=\f{1}{2} \pm \sr{\f{1}{4} - \f{m^2\ell^2 - 2\n m\ell}{3-\n^2} -\f{3(1+\n^2)\td k^2}{4\n^2} }
\eea

The boundary-boundary thermal correlators of the vector could be defined in a similar way as the scalar \cite{Chen:2009cg}. The frequencies of the vector quasinormal modes are the poles of the correlators in the momentum space.

\subsection{Tensor perturbation}

From the equation for tensor perturbation
\be
\e_\l^{\phantom{\l}\m\n} \na_\m h_{\n\s} \pm m h_{\l\s}=0,
\ee
which leads to
\be
\td \D h_{\phi\phi}= \lt( m^2 \pm \f{4\n m}{\ell}+\f{3\n^2}{\ell^2} \rt) h_{\phi\phi}.
\ee
To get the quasinormal modes of the tensor perturbation we just make the change $m^2 \ra m^2 \pm \f{4\n m}{\ell}+\f{3\n^2}{\ell^2}$ in the scalar results, and we get
\bea \label{e14}
&&\ \o_R=-i 2\pi T_R (n+h_R^\pm), \nn\\
&&h_R^\pm=\f{1}{2} \pm \sr{\f{1}{4} - \f{m^2\ell^2 + 4\n m\ell+3\n^2}{3-\n^2} -\f{3(1+\n^2)\td k^2}{4\n^2} },
~~ \textrm{or} \nn\\
&&h_R^\pm=\f{1}{2} \pm \sr{\f{1}{4} - \f{m^2\ell^2 - 4\n m\ell+3\n^2}{3-\n^2} -\f{3(1+\n^2)\td k^2}{4\n^2} }.
\eea
Similarly, these frequencies are the poles in the corresponding boundary-boundary thermal correlators.

\subsection{Spinor perturbation}

To solve the Dirac equations in the self-dual warped dS$_3$ spacetime, we should choose the vielbein for the background and compute the corresponding spin connection. Similar to the treatment of warped dS$_3$ black hole, we parameterize the metric of self-dual warped dS$_3$ spacetime in the form
\bea
ds^2=\f{gg^{rr}+g_{t\phi}^2}{g_{\phi\phi}} dt^2+\f{dr^2}{g^{rr}}+2g_{t\phi}dtd\phi+g_{\phi\phi}d\phi^2,
\eea
where we have two functions
\be
g^{rr}=\f{(r-r_b)(r_c-r)}{-A_1}, ~~~ g_{t\phi}=A_2 r+A_3,
\ee
and five constants
\bea
&&g_{\phi\phi}=\f{4\a^2\n^2\ell^2}{(3-\n^2)^2}, ~~~ g=-\f{4\a^2\n^2\ell^6}{(3-\n^2)^4},  \nn\\
&& A_1=-\f{\ell^2}{3-\n^2}, ~~~ A_2=\f{4\a\n^2\ell^2}{(3-\n^2)^2},
~~~ A_3=-\f{2\a\n^2\ell^2(r_b+r_c)}{(3-\n^2)^2}.
\eea
The vielbeins $e_\m^a$ are chosen as
\be
e^0=\sr{\f{-g g^{rr}}{g_{\phi\phi}}}dt, ~~~ e^1=\f{dr}{\sr{g^{rr}}},
 ~~~ e^2=\sr{g_{\phi\phi}}d\phi+\f{g_{t\phi}}{\sr{g_{\phi\phi}}}dt,
\ee
where $e^a=e^a_\m dx^\m$. The spin connection can be calculated straightforwardly and the nonvanishing components of the spin connection are
\bea
&& \o_t^{01}=-\o_t^{10}=-\f{g g'^{rr}+g_{t\phi}g'_{t\phi}}{2\sr{-g g_{tt}}},  ~~~
\o_\phi^{01}=-\o_\phi^{10}=-\f{g'_{t\phi}}{2}\sr{\f{g_{\phi\phi}}{-g}}, \nn\\
&& \o^{02}_r=-\o_r^{20}=-\f{g'_{t\phi}}{2\sr{-g g^{rr}}}, ~~~
\o^{12}_t=-\o_t^{21}=-\f{g'_{t\phi}}{2}\sr{\f{g^{rr}}{g_{\phi\phi}}}.
\eea

The Dirac equation is
\be
\g^a e^\m_a \lt( \p_\m +\f{1}{2}\o_\m^{ab}\S_{ab} \rt)\Psi + m\Psi=0,
\ee
where $\S_{ab}=\f{1}{4}[\g_a,\g_b], ~ \g^0=i\s^2, ~ \g^1=\s^1, ~ \g^2=\s^3$. With the ansatz $\Psi=(\psi_+,\psi_-)e^{-i\o t+ik\phi}$, we have
\be
\sr{g^{rr}}\p_r \psi_\pm +\f{g'^{rr}}{4\sr{g^{rr}}}\psi_\pm
   \pm \f{i(g_{\phi\phi}\o+g_{t\phi}k)}{\sr{-g g_{\phi\phi} g^{rr}}}\psi_\pm
   +\lt( m-\f{g'_{t\phi}}{4\sr{-g}} \mp \f{ik}{\sr{g_{tt}}} \rt)\psi_\mp=0,
\ee
The equations can be transformed into the form
\bea
&& \p_r(r-r_b)(r_c-r)\p_r \psi_\pm
+\f{\lt[ 4A_1\lt( g_{\phi\phi} \o + (A_3+A_2 r_b)k \rt) \pm i\sr{-g g_{\phi\phi}}(r_c-r_b) \rt]^2}
   {-16gg_{\phi\phi}(r_c-r_b)(r-r_b)}\psi_\pm    \nn\\
&&+\f{\lt[ 4A_1\lt( g_{\phi\phi} \o + (A_3+A_2 r_c)k \rt) \mp i\sr{-g g_{\phi\phi}}(r_c-r_b) \rt]^2}
   {-16gg_{\phi\phi}(r_c-r_b)(r-r_b)}\psi_\pm    \nn\\
&&=\lt( \f{1}{4} -A_1(m-\f{A_2}{4\sr{-g}})^2 +\f{A_1(A_1 A_2^2+g)k^2}{-g g_{\phi\phi}} \rt)\psi_\pm, \nn
\eea
which is just
\bea
&& \p_r(r-r_b)(r_c-r)\p_r \psi_\pm
+\f{\lt[ 4\lt( \o-\f{r_c-r_b}{2}\f{k}{\a} \rt) \mp i(r_c-r_b) \rt]^2}{16(r_c-r_b)(r-r_b)}\psi_\pm   \nn\\
&&+\f{\lt[ 4\lt( \o+\f{r_c-r_b}{2}\f{k}{\a} \rt) \pm i(r_c-r_b) \rt]^2}{16(r_c-r_b)(r_c-r)}\psi_\pm
=\lt( \f{1}{4} +\f{(m\ell-\n/2)^2}{3-\n^2} +\f{3(1+\n^2)k^2}{4\a^2\n^2} \rt)\psi_\pm. \nn
\eea

In the region $r_b<r<r_c$, we should impose the outgoing boundary conditions at two cosmological horizons. At the horizon $r=r_b$, the condition restricts
\bea
&& \psi_+ \pp z^{\a_f}(1-z)^{\b_f}F(a_f,b_f,c_f;z) , \nn\\
&& \psi_- \pp z^{\a_f-1/2}(1-z)^{\b_f}F(a_f-1,b_f,c_f-1;z),
\eea
with the new coordinate $z$ being
\be
z=-\f{r-r_b}{r_c-r},
\ee
and the parameters
\bea
&&\a_f=-i\f{2\a\o-(r_c-r_b)k}{2\a(r_c-r_b)} -\f{1}{4},    \nn\\
&&\b_f=\f{1}{2}+\sr{ -\f{(m\ell-\n/2)^2}{3-\n^2} -\f{3(1+\n^2)k^2}{4\a^2\n^2} },  \nn\\
&&\g_f=-i\f{2\a\o+(r_c-r_b)k}{2\a(r_c-r_b)} +\f{1}{4},  \nn\\
&&a_f=\a_f+\b_f-\g_f=\b_f-\f{1}{2}+i\f{k}{\a},  \nn\\
&&b_f=\a_f+\b_f+\g_f=\b_f-i\f{2\o}{r_c-r_b},  \nn\\
&&c_f=1+2\a_f.
\eea
  The outgoing boundary condition at the cosmological horizon $r=r_c$ leads to the constraint
\be
b_f=-n, ~~ \textrm{or} ~~ c_f-a_f=-n,
\ee
from which we can read the quasinormal modes of the spinor perturbation in the self-dual warped dS$_3$ spacetime
\bea
&&\o_R=-i 2\pi T_R (n+h_R^\pm), \nn\\
&&h_R^\pm=\f{1}{2}\pm\sr{-\f{(m\ell-\n/2)^2}{3-\n^2} -\f{3(1+\n^2)\td k^2}{4\n^2}}.
\eea

\subsection{Algebraic construction}

In this subsection we construct the scalar, vector, and tensor quasinormal modes of the warped dS$_3$ spacetime from the hidden conformal symmetry in an algebraic way based on the formalism proposed in \cite{Chen:2010ik,Chen:2010sn,{Chen:2010fr}}, where the readers can see more details.

Under the background of a warped spacetime, the equation of motion of scalar, and some linear combinations of the vector or tensor components $\Phi$ can be written in the form
\be
\lt( \ma L^2 +K \rt) \Phi=\lt( \ma L^2 +b\ma L^2_{\td V_0} +a \rt) \Phi=0,
\ee
where $\ma L$ denotes the Lie-derivative and $\ma L^2$ denotes the Lie-induced quadratic Casimir
\be
\ma L^2 \equiv -\ma L_{V_0} \ma L_{V_0} +\f{1}{2} \lt( \ma L_{V_1}\ma L_{V_{-1}}
                                                       + \ma L_{V_{-1}}\ma L_{V_1} \rt),
\ee
and $K$, $a$ and $b$ are some constants to be fixed.

As the construction of various kinds of quasinormal modes is similar, here we take the scalar field as the example to illustrate the method. Firstly, we define the highest weight state as
\be \ma L_{V_0}\Phi^{(0)}=h_R \Phi^{(0)}, ~~~ \ma L_{V_1}\Phi^{(0)}=0, \ee
and then, from the highest weight state we construct the descendent modes
\be \Phi^{(n)}=\ma L^n_{V_{-1}}\Phi^{(0)}. \ee
We also define $\ma L_{\td V_0}\Phi=q\Phi$, then we have
\be K=b q^2+a. \ee
Since the Lie-induced Casimir $\ma L^2$ always commute with the Lie-derivatives, with $\ma L^2$ operating on the descendent modes $\Phi^{(n)}$, we have
\be h_R^2-h_R-K=0, \ee
and then we get the conformal weight
\be
h_R=\f{1}{2} \pm \sr{\f{1}{4}+K}.
\ee
To compute the frequencies of the quasinormal modes, we expand $\Phi=e^{-i\o t+ik\phi}$. With $\ma L_{V_0}$ operating on the descendent modes $\Phi^{(n)}$ we get the frequencies
\be
\o=-\f{n_L}{\pi T_L}k+i\f{2A}{T_L}(n+h_R).
\ee
Note that only when $A<0$, the above frequencies have the right behavior of quasinormal modes. However in the case of the self-dual warped dS$_3$ spacetime, we have $A>0$. We have to redefine the $SL(2,R)$ generators as what was done in \cite{Chen:2010sn} by setting
\be
\hat V_1=V_{-1}, ~~~ \hat V_0=-V_0, ~~~ \hat V_{-1}=V_1.
\ee
With the new $SL(2,R)$ generators we can get frequencies of the quasinormal modes
\be
\o=-\f{n_L}{\pi T_L}k-i\f{2A}{T_L}(n+h_R).
\ee
We can solve the highest wight condition $\ma L_{\hat V_1}\Phi^{(0)}=0$ as
\be
R^{(0)}=C (r-r_b)^{-i\f{(\pi T_L-n_R)\o-(\pi T_R-n_L)k}{4A}}
          (r_c-r)^{-i\f{(\pi T_L+n_R)\o+(\pi T_R+n_L)k}{4A}},
\ee
where $C$ is a constant. When $r \ra r_b+0$, we have
\be
\Phi^{(0)} \sim e^{-i\o \lt( t+\f{(\pi T_L-n_R)}{4A}\ln(r-r_b) \rt)},
\ee
which just has the outgoing boundary condition. Also when $r \ra r_c-0$, we have
\be
\Phi^{(0)} \sim e^{-i\o \lt( t+\f{(\pi T_L+n_R)}{4 A}\ln(r_c-r) \rt)},
\ee
which just has the outgoing boundary condition. We can see $\Phi^{(0)}$, and thus all $\Phi^{(n)}$ have the right behavior as the quasinormal modes.

In the case of self-dual warped dS$_3$ spacetime, we have
\bea
&&b=\f{3(1+\n^2)}{4\n^2}, ~~~ q=i\f{k}{\a}, ~~~ a=-\f{m^2\ell^2}{3-\n^2},  \nn\\
&&K=-\f{m^2\ell^2}{3-\n^2}-\f{3(1+\n^2)k^2}{4\a^2\n^2}.
\eea
We get the same quasinormal modes (\ref{z1}) for the scalar perturbation. For the vector and tensor perturbations the quasinormal modes can be constructed similarly. Unlike $\td V_0$, the vector $V_0$ has vanishing $\phi$ but nonvanishing $t$ components, so we can not use another set of vectors $(\td V_0, \td V_{\pm1})$ to construct the quasinormal modes.

\section{Conclusion and discussion}

In this paper we studied several properties of the self-dual warped dS$_3$ spacetime. There are two cosmological horizons in this spacetime. We investigated the thermodynamics at both horizons, and found that they are more or less the same. We conjectured that the self-dual warped dS$_3$ spacetime can be described by a two-dimensional finite temperature chiral CFT. We analyzed the scalar wave function carefully and got the boundary-boundary correlators at the future timelike infinity. We also got the quasinormal modes of various perturbations of the spacetime. The frequencies of various kinds of quasinormal modes take the same form
\be
\o_R=-i 2\pi T_R (n+h_R),
\ee
which correspond to the poles in the boundary-boundary thermal correlators in the momentum space. This is a strong evidence for the conjectured warped dS/CFT correspondence. Moreover we investigated the hidden conformal symmetry of the self-dual warped dS$_3$ spacetime in the physical region and constructed the towers of the quasinormal modes with the help of the ladder operators, which are in perfect match with what were obtained by solving the equations of motion.

One central issue in warped dS/CFT correspondence is that there is short of derivation of the central charge (\ref{central2}) from asymptotic symmetry analysis. It seems that there is ambiguity in the formalism developed in \cite{Compere:2008cv}.\footnote{We would like to thank S.~Detournay for pointing out this subtlety.} This ambiguity hinders us from determining the central charges. Note that the left-moving central charge comes from the enhancement of U(1) isometry along the rotation in $\phi$ direction. Moreover, it should be emphasized that the possible boundary conditions for the self-dual warped dS$_3$ spacetime leading to the chiral CFT must be different from the ones for the de Sitter-esque geometry discussed in \cite{Anninos:2011vd}, even though both geometries may share the same asymptotic geometry. This present another example that in 3D TMG gravity, different boundary conditions may lead to different dual CFTs, as happened in warped AdS$_3$ case\cite{Chen:2010qm,{Compere:2009zj}}.

On the other hand, it is also remarkable that for the self-dual warped dS$_3$ spacetime, there is no right-moving sector. Recall that the dS$_3$ spacetime itself could be written as the U(1) bundle over a dS$_2$ as well. In TMG theory, dS$_3$ spacetime should be dual to CFT with both sectors, one from the U(1) enhancement, the other from SL(2,R) of dS$_2$.  Therefore the absence of the right-moving sector in self-dual warped dS$_3$ case is quite surprising and deserves further study.

\vspace*{10mm}

\noindent
{\large{\bf Acknowledgments}}

The work was in part supported by NSFC Grant No. 10975005. We thank S.~Detournay and Bo Ning for valuable correspondence and discussions.

\end{document}